\documentclass[%
 reprint,
 amsmath,amssymb,
 aps,
]{revtex4-1}

\usepackage{graphicx}% Include figure files
\usepackage{dcolumn}% Align table columns on decimal point
\usepackage{bm}% bold math
\begin{document}

\preprint{APS/123-QED}

\title{In-depth analysis of music structure  as a text network}% Force line breaks with \\

\author{Ping-Rui Tsai$^{1}$, Yen-Ting Chou$^{1}$, Nathan-Christopher Wang$^{2}$, Hui-Ling Chen$^{3}$, Hong-Yue Huang$^{1}$,  Zih-Jia Luo$^{4}$, and Tzay-Ming Hong$^{1\ast}$}
\affiliation{%
$^1$Department of Physics, National Tsing Hua University, 
Hsinchu 30013, Taiwan, R.O.C}
\affiliation{%
$^2$College of Pharmacy, University of Michigan, Ann Arbor, MI 48109 U.S.A.}%
\affiliation{%
$^3$Department of Chinese Literature, National Tsing Hua University, 
Hsinchu 30013, Taiwan, R.O.C}
\affiliation{%
$^4$Advanced Semiconductor Engineering, INC., Kaohsiung 76027628, Taiwan, R.O.C}

\begin{abstract}
Music, enchanting and poetic, permeates every corner of human civilization. Although music is not unfamiliar to people, our understanding of its essence remains limited, and there is still no universally accepted scientific description. This is primarily due to music being regarded as a product of both reason and emotion, making it difficult to define. In this article, we focus on the fundamental elements of music and construct an evolutionary network from the perspective of music as a natural language, aligning with the statistical characteristics of texts. Through this approach, we aim to comprehend the structural differences in music across different periods, enabling a more scientific exploration of music. Relying on the advantages of structuralism, we can concentrate on the relationships and order between the physical elements of music, rather than getting entangled in the blurred boundaries of science and philosophy. The scientific framework we present not only conforms to past conclusions in music, but also serves as a bridge that connects music to natural language processing and knowledge graphs.
 
%\begin{description}
%%Secondary publications and information retrieval purposes.
%May be entered usintog the \verb+\pacs{#1}+ command.

%\end{description}
\end{abstract}

\pacs{Valid PACS appear here}% PACS, the Physics and Astronomy
\maketitle

\section{Introduction}

Music is often considered as a form of natural language \cite{Music_Nature1,Music_Nature3} because it exhibits the ability to adapt \cite{Nature1} and co-evolve with the human civilization. According to Soviet musicologist Henry Orlov, music can participate in communication and unite people through a single emotion. 
The development of languages and music can be categorized into distinct stages that reflect the historic events of each period, and both originate from the imitation of sounds from the environment. 
Music, like language, has developed its own notation and system of reading and writing. The meaning conveyed by music is less precise than spoken language due to its lack of tenor and vehicle \cite{tenor1,tenor2}. Despite this ostensible weakness, music is no less capable of evoking our memories of specific experiences \cite{music_memory}.
By using magnetoencephalography (MEG), it has been confirmed that music and language are governed by the same mechanism in the cerebral cortices.  This implies the similarity in their processing of data that are transmitted through the sounds associated with spoken language and music harmony \cite{music_text1}. 

\begin{figure}
\centering
\includegraphics[width=8cm]{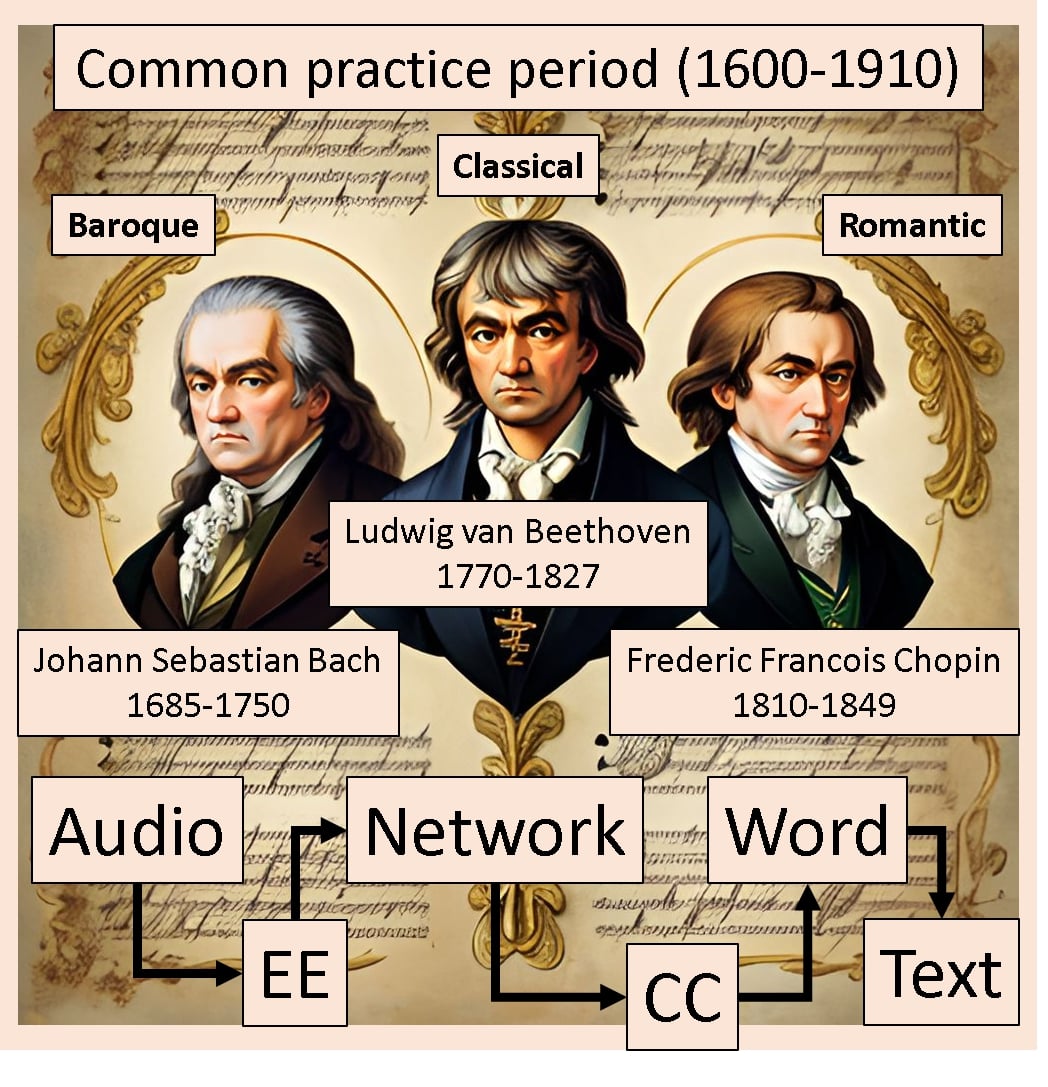}\caption{This illustration depicts the three main periods we primarily discuss in the article: the Baroque, Classical, and Romantic periods, each represented by a representative musician. Additionally, we visually present the overall processes of the textualization of music in EE.}
\end{figure}  %Although language is rarely expressed through singing or playing an instrument like music, they both make use of the same cognitive mechanism which gives rise to the active discipline of ``Shared Synthetic Integration Resource Hypothesis" \cite{music_text2}.

Let's proceed to compare their structure. Both involve minimal units, such as words and chunking, that serve as the building blocks to construct the corpus and scores through ever larger hierarchical units, i.e., phrases/sentences and the  deep structure \cite{structure1,structure2}. It is equally interesting to look into the correlation within the same unit and across different units. 
With the development of statistical linguistics, Zipf \cite{Zipfs1,Zipfs2} empirically found that the frequency-rank distribution of corpus and natural language utterances follows the power law: $y=a/x^{b}$ where $a,b$ are constants. This distribution was later established to be prevalent in the ranking of many natural and man-made systems, such as web link \cite{Power1} and brain functional network \cite{Power2,Power3}. 
Although Zipf's law has been confirmed to exist in musical composition, there is no consensus on what unit plays the role of word. For example, it has been proposed via segmentation aided by the application of equal temperament to pitch, timbre, and loudness or through binary coding on the power spectrum \cite{zipfs_music1,zipfs_music2}. The numerous attempts above, regardless of the way word definitions are approached, all revolve around definitions based on physical categories. % C1

In this article, we deconstruct fundamental musical elements like rhythm, timbre, pitch, melody, articulation, meter, and tempo, and then transform them into concepts derived from physics — specifically, space, time, and volume. These three elements are herein referred to as essential elements (EE). The notes ($C^{1}, D^{1}, E^{1}, E^{1\#}, G^{2}, F^{2}$, etc.) and time, along with its corresponding beat, are important representations of the universal features of music \cite{UF}. Employing a reference coordinate system based on 0.1-second intervals and piano keys, we establish linking conditions between each pixel (node) to form an evolutionary network. Additionally, we define words by clustering coefficients (CC) in order to transform musical audio into texts, see Fig. 1.

Here we will answer five questions: (1) How to generate text from music using all of the essential elements in music?
(2) What are the linking conditions of music across different periods in an evolutionary network and the variations under Zipf's law distribution?
(3) How does the diversity in the choice and frequency of words within an evolutionary network across different periods reflect the versatility in song structures?
(4) How robust and how free are the network that reflects the characteristic features of each music periods against random removal of words?
(5) How do we use a structure of audio to discern music from non-music? And how can we observe the evolution and extinction of musical words akin to the evolution of natural language by tracking the magnitude of CC?

\section{ network modeling}

To answer the first question that is inspired by previous methodologies in network science for generating specific network structures \cite{BR1}, we establish linking conditions based on interactions among EE. Subsequently, these conditions can be used to  determine the network structure. We quantify the variations between these EE and establish a threshold  for the validation of links to form a  structure. The music signals can then be analyzed in the frequency and time-frequency domains, with a frequency range of 1$\sim$8192 Hz and a time increment of 0.1 second. Additionally, we will transform these domains into the note-time domain. The note consists of 84 keys, which correspond to those of a piano in equal temperament and cover a frequency range of 1$\sim$8192 Hz. Expressing the volume in normalized decibels from 0 to 10, we eliminate the volume that is less than 0.1, based on the power-time spectrum in order to reduce the difference in volume that was derived from the recording process. 

\begin{figure}
\centering
\includegraphics[width=8cm]{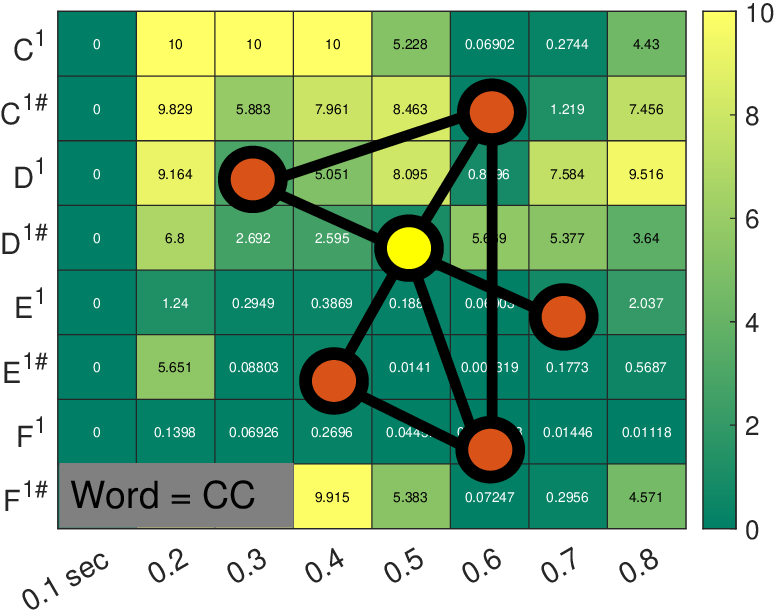}\caption{Volumes are stated in the note vs. time plot for the music score by Ryuichi Sakamoto.  The exchange of information between two pixels is determined by their elements and  weights. Using Eq. (1), we calculate the linking condition for the text network of the music and define the link between orange and yellow dots. The CCs in the network are then treated as words.}
\end{figure}  

We then define the amount of information (I) via comparing the change of note position (N), time position (T), and volume (V) of pixel 1 and 2 in note vs. time coordinate: 
 \begin{equation}
 \begin{split}
 I \equiv &w_1\cdot|{\rm N}(1)-{\rm N}(2)|+w_2\cdot|{\rm T}(1)-{\rm T}(2)|\\
 +&w_3\cdot|{\rm V}(1)-{\rm V}(2)|+w_4\cdot|{\rm V}(1)+{\rm V}(2)|
 \label{information}
 \end{split}
 \end{equation}
  How these  weights  $w_{1,2,3,4}$ are chosen will be explained shortly. Via the simple linear regression in Eq. (1), the weights can be determined and shown to offer a high level of interpretability of the musical form  across these four aspects when composing. The first three terms in Eq. (\ref{information}) reflect the signature of composition, while the fourth is an additional term that represents the energy carried by the positions and emotions that the composer tried to convey \cite{Emotion_dynamic}.
 The amount of information carried by any music is limited and determined by both the composer's style and motives \cite{M1}. We rarely see dramatic changes in all elements in musical pieces and, therefore, it is expected that there exists an upper threshold $I_{\rm m}$ for the amount of information that can be conveyed between pixels in the local interaction:

\[ Link =
\begin{cases}
1 & \text{if } I < I_{\rm m},\\
0 & \text{if } I > I_{\rm m}.
\end{cases} \]

This also implies that, within limited variations, the priorities considered by the composer determine the style of composition. % 20231212

  Using the total number of links, we can calculate ${\rm CC}=2N/[K(K-1)]$ of each pixel, where the degree $N$ counts the number of neighbors and $K$ the number of links among them. In order for CC to be interpreted as a word, it has to be significant enough at supplying the semantic meaning of the cluster based on the quantities $N$ and $K$. A sample results of the above procedures can be found in Fig. 2.
This definition of word not only is consistent with the structuralism of musical material \cite{Syntax1}, but also follows the spirit of natural language modeling\cite{text_mining1,text_mining2}, such as linguist John Rupert's interpretation of ``You shall know a word by the company it keeps"\cite{John}. This spirit expresses the role of endowing meaning to a word via its interaction with neighbors. 

Having defined the equivalent of words in music is not enough. We still need to make sure that the structure derived from them obeys the statistical properties in linguistics, notably the empirical Zipf's law. This imposes a constraint that can be utilized to select the most suitable weights $w$ and $I_{\rm m}$ and threshold that give the best fitting by a power-law distribution for  word frequency vs ranking. Finally, the  $w$ and $I_{\rm m}$ function like a fingerprint  that is unique to the structure and helpful for us to distinguish the music from different periods. We can also use them to compare the properties of musical and non-musical  structures, as the composer Edgar Var\'ese said, ``Music is organized sound" \cite{Edgar}.

We need to select the optimal  weight combination from 4032 configurations to accurately represent the text representation of the music.  weights set the range for note and time weights to be 0.05 $\sim$ 0.3 with an interval of 0.05 and the range for volume to be 0.1 $\sim$ 0.4 with an interval of 0.1. We also normalized the decibel values between 0 and 10 for all data points. The threshold range was set to be 0.5 $\sim$ 1.8 with an interval of 0.2. In order to reduce the computing time, we will focus only on the neighboring nodes that are separated within 7 notes and 0.7 sec.  

The optimal values for the  weights in Eq. (\ref{information})  were determined by two criteria: first, the distribution of Zipf's law for CC in the music score must exhibit an R-square value exceeding 0.8 after deleting the first rank and plotting the frequency vs  rank in full logarithm. Second, the largest type of CC should be selected as the optimization condition in order to extract the maximum number of word types to maintain diversity. 

From now on, we shall name our  evolutionary network as the Essential Element Network (EEN) to emphasize its inclusion of essential EE in music. Our analyses focus on piano pieces from the Common Practice Period (CPP) which spans from 1650 to 1900 AD\cite{CPP}. The CPP marks the establishment of Western musical system and the definition of harmony system which is essential for the interaction between musical elements \cite{IN1,IN2}. Within the CPP, the Baroque period (B) is the earliest, followed by the period of classical music such as Beethoven (BN), and the Romantic period (R) is the most recent.
% ok 20221214

\begin{figure}
\centering
\includegraphics[width=9cm]{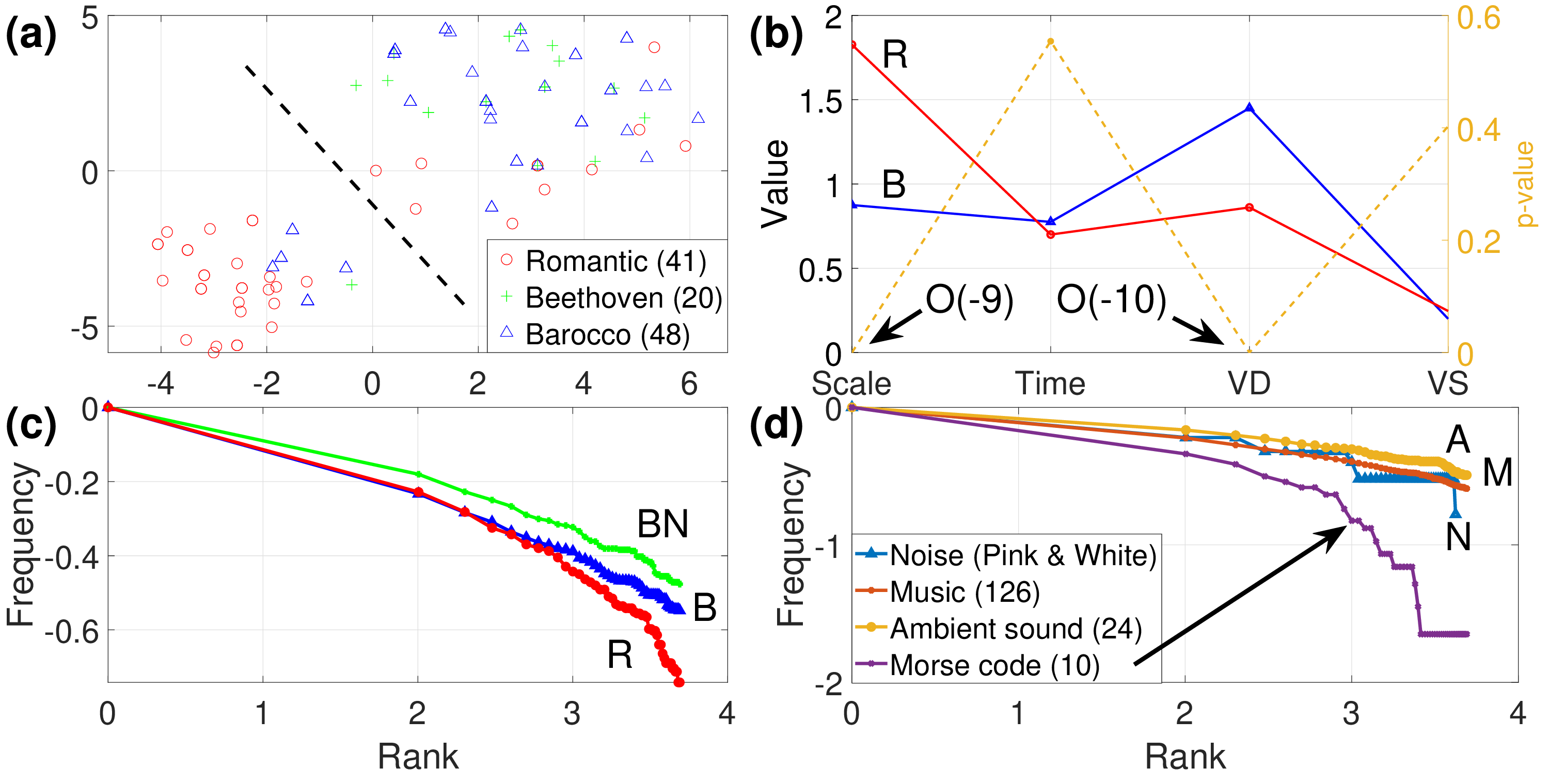}\caption{(a) A t-SNE mapping of the four weights and threshold value onto the eigenspace. The dash line is to highlight the existence of two clusters, as indicated by the statistical population in parentheses. (b) A t-test is conducted to assess the statistical significance of the weight selection where the orange dotted line is for p-value on the right y-axis.  (c) This full logarithmic plot for the Zipf distribution in different periods. (d) The Zipf distribution  of different types of sounds under the range of weight selection, We use the initial letter to represent a line, in which ambient sound includes bird, river, and city traffic.} 
\end{figure} 

\section{Distinguishing Different Musical Periods}
\subsection{Criterion: Weights}
To answer  question (2), we can examine the weights of music from different periods to understand the variations in network structure. We used the t-distributed stochastic neighbor embedding (t-SNE) as a dimension reduction method to visualize the distribution of weights in three different musical periods \cite{TSN1}. The clustering of data points for music from the Beethoven and Baroque periods in Fig. 3(a) implies their weights are more similar to each other than to the Romantic. In Fig. 3(b), it can be seen that during the Baroque period, $w_3$ is the most prominent weight, which is related to the performance of clear gaps between notes, with the aim of maintaining clear voices. 
 
 The $w_1$ and $w_2$ are chosen with similar weights, representing the characteristics of Baroque polyphonic music, emphasizing rigor counterpoint\cite{ABS_M}. However, the priority of weights shifts in the Romantic period when people tried to break free from the constraints of composition, leading to a greater freedom and evolution of more diverse composition styles \cite{RM1, RM2}. A t-test analysis of the weights shows that there exists a significant distinction between $w_1$ and $w_3$ if the p-value is less than 0.001. The combination of weights behind the EEN reflects the characteristic style of composition in each period by use of Zipf's law, as shown in Fig. 3(c).  In Fig. 3(d), we also compare different types of audio to test the applicability in 4032 combinations of weights. It turns out that  Morse code did not follow the expected power-law distribution, which suggests that the chosen weights are not suitable.
\begin{figure}
\centering
\includegraphics[width=9cm]{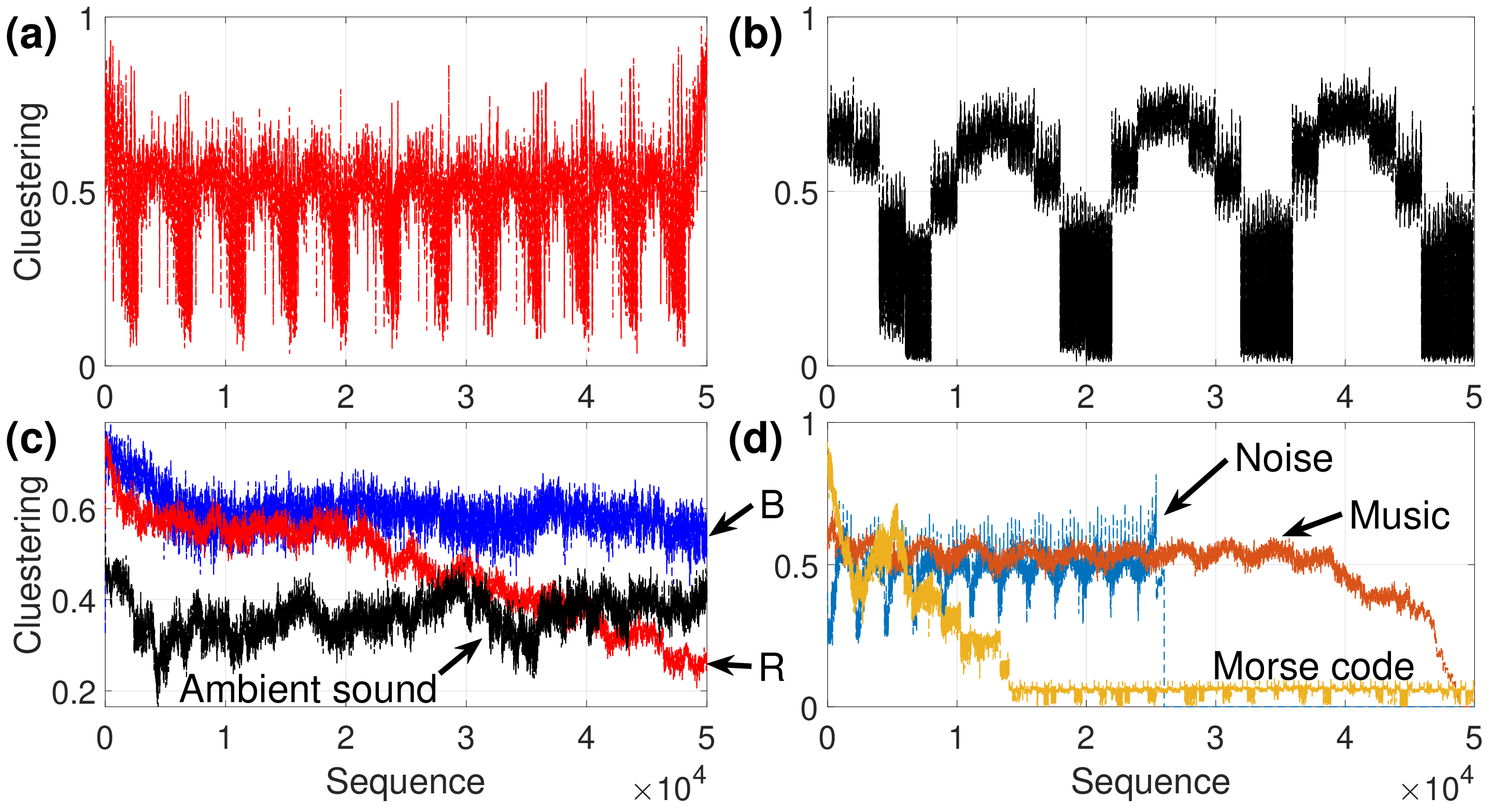}
\caption{The CC is plotted against sequence. Two samples of EEN distributions in one dimension are shown in (a, b), both of which turn out to be periodic. (c) Compared to the Baroque period and ambient sound, the variation of CC in the Romantic period is more pronounced which is in line with the conventional view in musical history that its musical form is more diverse. (d) The distribution of different types of audios where the music samples include 38 piano songs with a length of less than two minutes} 
\end{figure}  

\subsection{Criterion: Trend and Histogram of Words}

Here we will answer question (3). After CC is calculated by scanning from the first to last notes, the resulting sequence is found to be periodic  in Fig. 4(a, b). Based on the sequence, we can analyze the representation of words. Figure 4(c) shows that the Baroque period, which emphasized musical formalism, has a uniform development in performance. In contrast, the Romantic period adopted a strategy of destroying or escaping the previous musical form. Ambient sound is shown to share a similar distribution to the Baroque period, but with more fluctuations and different CC intervals. In Fig. 4(d), we found that the distribution of CC for music resembled that of white and pink noises, in contrast to the of Morse code. This is reasonable because music often contains elements that are abstract and hard to be assigned any meaning, whileas the tenor and vehicle of Morse code are always precise.

Having shown the trend of CC, let's now analyze their histogram in Fig. 5. The normal distribution is obtained for the ambient sound as well as all musical periods, except the Romantic one which exhibits multiple peaks which reflect  more diversity in its selection of words.
 
\begin{figure}
\centering
\includegraphics[width=9cm]{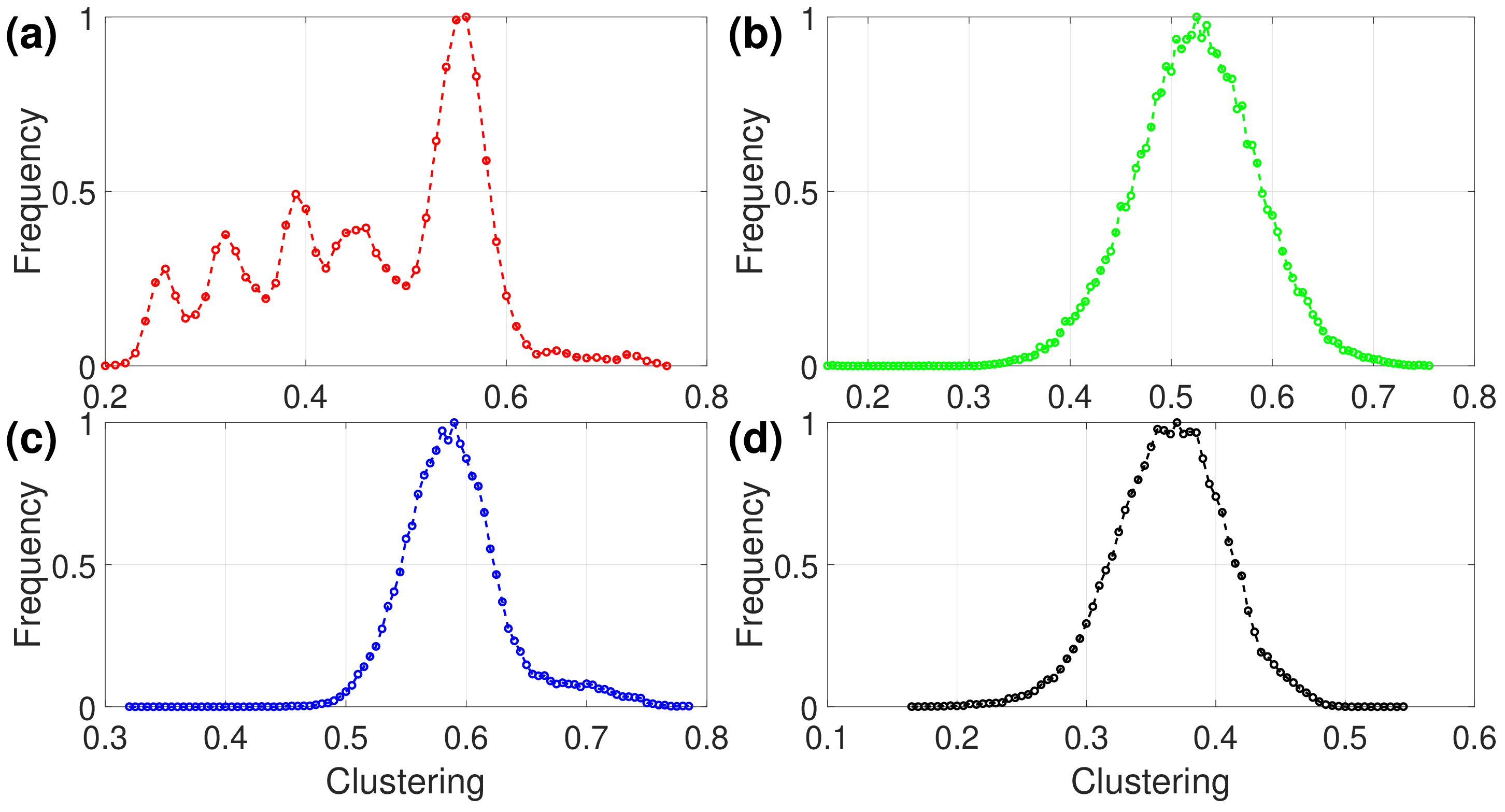}
\caption{The histogram of Romantic, Beethoven and Baroque period, and the  ambient sounds is shown respectively in (a), (b), (c), and (d). The Romantic period is the only one that defies the normal or Gaussian distribution at exhibiting multiple peaks.
} 
\end{figure}  

\section{Deep Learning analyses}
\subsection{Training Information}

\begin{figure}
\centering
\includegraphics[width=9cm]{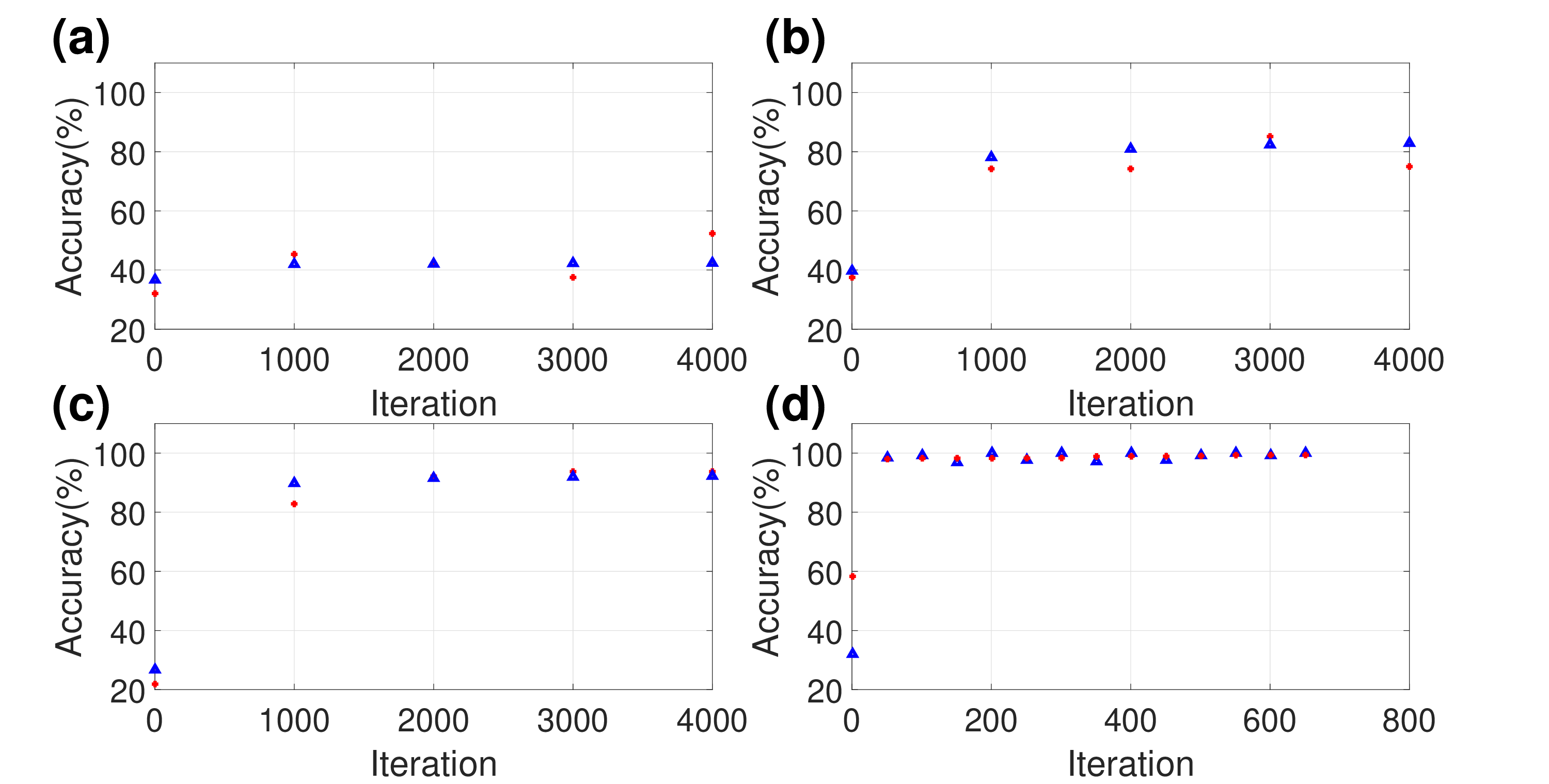}
\caption{(a), (b), and (c) represent the learning curves for minimum text size to  classify three CPP periods for texts with (2 notes, 0.3 sec), (72 notes, 3.5 sec), and (84 notes, 10.1 sec) where the blue/red line denotes the training/validation curve. (d) shows similar curves for classifying music or non-music with (84 notes, 10.1 sec).} 
\end{figure} 

To identify the unique characteristics of CPP, we use word mapping to represent CPP by a 2D EEN. This is done by filling the CCs to their corresponding pixel in the note vs. time plot of Fig. 2, and applying a convolutional neural network (CNN) \cite{CNN1} to classify musical periods. CNN is a powerful image classification model that can predict the label of an image, relying on its ability to extract local spatial features and keep translation invariant. 

To test the robustness and minimum discernible information carried by a text, we design the following two tasks: (1) determine the minimum size of text to represent each of the three CPP periods and (2) test the rigor of conventional definition for these periods. Treating the text as a network, we follow the conventional approach of  random node removal to explore the robustness of network feature. Instead of randomly deleting network nodes \cite{RRM1}, we remove words from 2D EEN to 
understand the maximum amount of disruption each CPP can withstand and  be effectively recognized by a CNN. 
We expect this knowledge can be used to infer the relative amount of rules in musical form. Strictness implies more rules that will likely render the network more vulnerable to disruptions. More details of the procedure and results of training shown respectively in Figs. 5 and 6 will be discussed in Sec. IV(B) and (C). 

Due to the numerous CNNs used in Fig. 7(a), we only provide the minimum and maximum amount of 2D EEN for the learning curves in Fig. 6(a) and (b) for task (1) with 2-notes/0.3-second and 72-notes/3.5-second from 34000 and 32456 samples. Similarly, we need to train the CNN to obtain a test accuracy exceeding 93\% for the prediction of 2D EEN's period in preparation for task (2). The result is shown in Fig. 5(c). In the meantime, Fig. 6(d) is dedicated to distinguishing music and non-music with a test accuracy equaling 100\%. Both Fig. 6(c, d) adopt 84-notes/10.1-second and are trained from 12,000 samples. So far, the CNN adopts four convolutional layers of 2$\times$2 kernel size and three fully connected layers with 128, 64, and 3 nodes. The hyperparameters are training-validation-test ratio of 6:2:2, batch size of 64, epoch of 6, and a learning rate of $10^{-4}$.

\subsection{Minimum Features and Robustness of Text Networks}

\begin{figure}
\centering
\includegraphics[width=9cm]{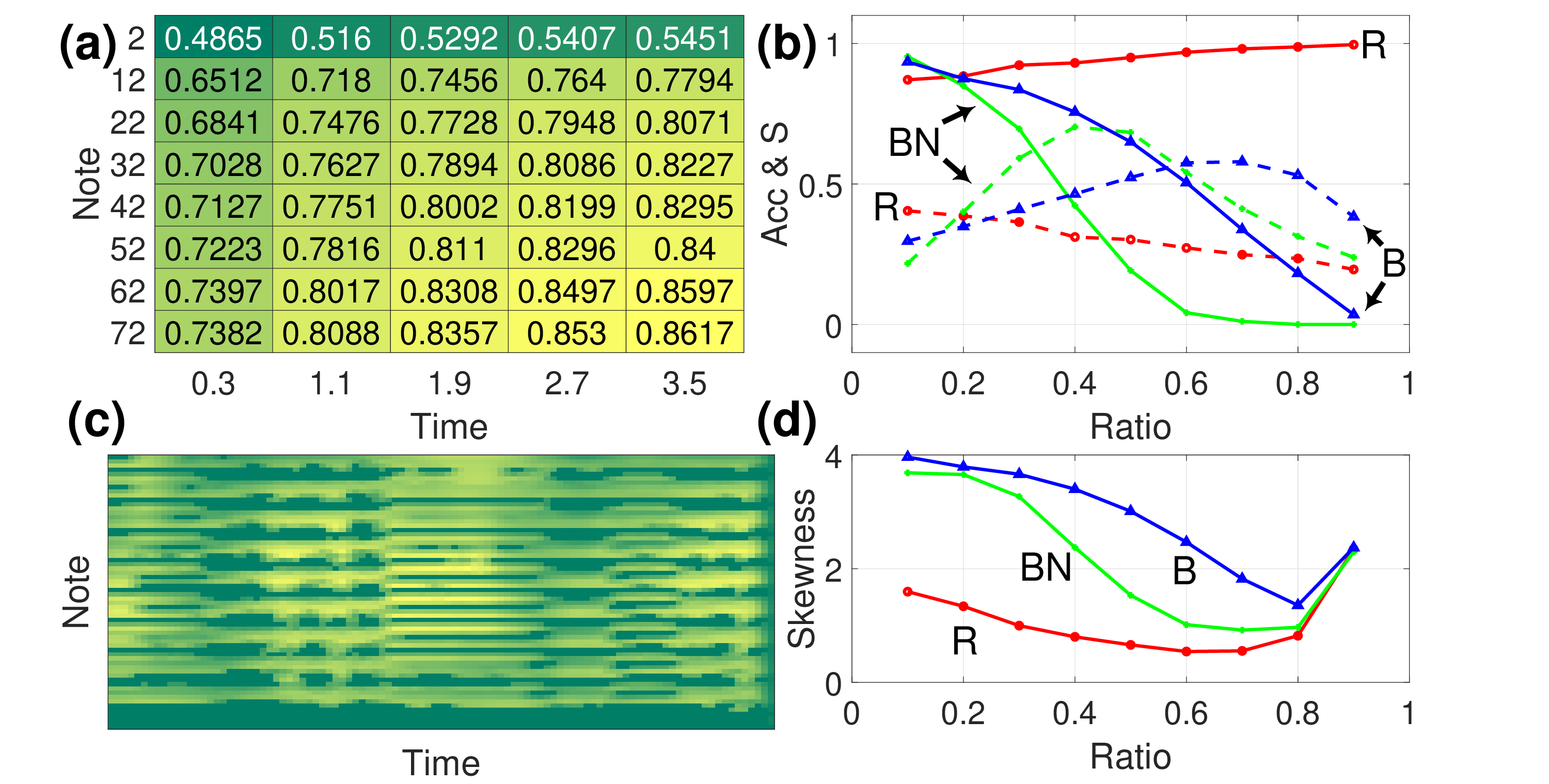}
\caption{(a) The accuracy of distinguishing Baroque, Beethoven, and Romantic periods is shown with different sizes of 2D EEN information, as defined in the text. (b) Following the label in Fig. 3(a), the dotted and solid lines represent Shannon entropy (S) and accuracy (ACC). The features of Romantic period are enhanced by discarding information. (c) The score distribution of 2D information is shown in grad CAM. Green color denotes a lower score. (d) Based on the magnitude of skewness, we can observe that the change in feature distribution due to the destruction of structure in the Romantic period is smaller than the other two periods.} 
\end{figure}  

In addition to answering how many words are required to distinguish musical periods, we are ready to address question (4). Figure 6(a) summarizes the accuracy for different sizes of 2D EEN which shows that, to obtain a  test accuracy exceeding 80\%, a minimum requirement is 42-note/1.9-second for task (1). As for task (2), we systematically increase the percentage of words removed from a text until a trained CNN changes its prediction. The Softmax layer in a CNN can convert features into probabilities for predicting labels \cite{SOFT}, which allow us to compute Shannon entropy for these three periods. By analyzing 1000 samples from each period, we find in Fig. 7(b) that the accuracy of the Baroque and Beethoven samples decrease with more deletions. Surprisingly, the accuracy of the Romantic samples improves. This can be interpreted from hindsight as that the CPP music before the Romantic period emphasized a more rigorous pursuit of musical form and, therefore, allows less tolerance for arbitrary disruptions \cite{RM1} in the coordination between different musical elements. In order to understand the characteristics of word distribution in a 2D EEN, we use grad CAM to extract the significant locations from the CNN \cite{GRA1}. Grad CAM, through the concept of back propagation, allows us to explore  the scoring criteria for labeling \cite{GRA2}, as shown in Fig. 7(c). 

In Fig. 7(d), we use skewness, a good measure of distributional asymmetry, to calculate and plot the score histogram by averaging over the samples after grad-CAM. We can observe that using skewness as a feature allows us to achieve temporal ordering between three periods based on the grad-CAM scoring criteria. It is also evident that, in the context of random word disruption, Beethoven's representation of the middle period tends to exhibit skewness closer to  Romantic period, while the decay trend in the Baroque period is comparatively gradual. The fact that the skewness is larger for the two periods before the Romantic implies the former did not have as many prominent features. This also means that the features were considered in a more holistic manner. Additionally, the score distribution during the Romantic period is closer to Gaussian. We can observe that, as the text structure gradually breaks down, indicating a departure from strict structural conventions, the less disrupted and more intact conditions closely resemble the Baroque period. Conversely, as the degree of disruption increases, the features tend to approach those of the Romantic period. The fact that this conclusion is reproduced in Fig. 7 vindicates the strength and accuracy of our approach.
%20221214 !!!
\begin{figure}
\centering
\includegraphics[width=8.5cm]{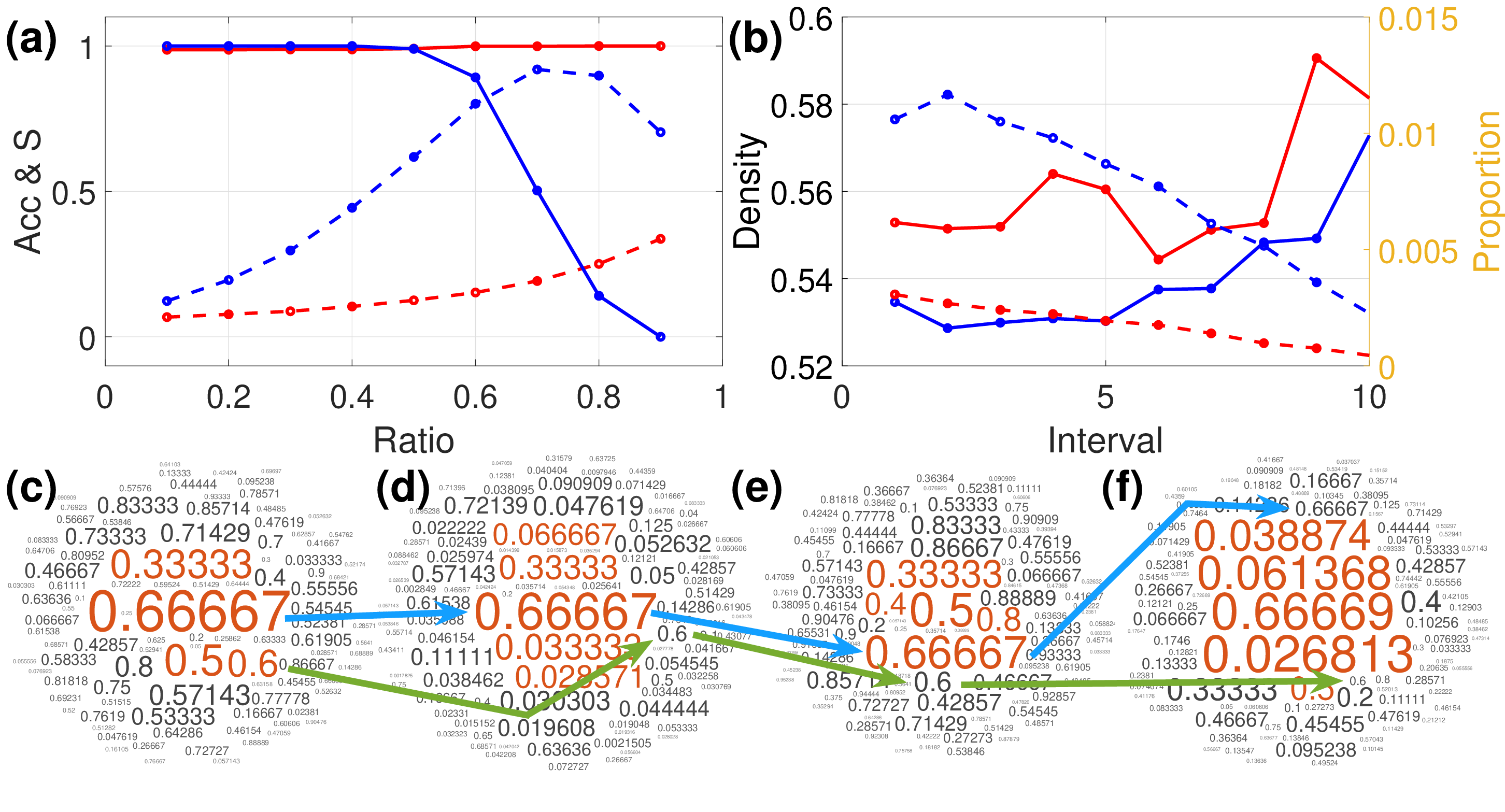}
\caption{In (a-b), the blue circle and red cross denote non-music and music. In (a), the solid and dashed lines represent accuracy (Acc) and Shannon entropy (S). (b) The Dashed/solid line follows the note of the right/left y axes. (c-f) show the term changes of Baroque, Beethoven and Romantic  periods, and modern Japanese composer's musics, such as those by Ryuichi Sakamoto, respectively. The destruction of terms is demonstrated by the termination of arrows, while new words can be seen to emerge.} 
\end{figure}  
%OF_5_GAN_grad_cam.eps

\subsection{Difference Between Music and Non-music and the Evolution of Words}
We attempt to address the difference between music and non-music through the fifth question.
It is widely believed that the origin of music can be traced back to ancient humans imitating natural and ambient sounds\cite{O1,O2}. To investigate this idea, we trained a network to distinguish between music and non-music  including  ambient sound and noises by 64 note and 10.1 sec. By the same process in Fig. 7(b), we detected the characteristics of 2D EEN in Fig. 8(a), and found that random destruction of the structure actually preserved its characteristics. We know that ambient sound and noise lack clear rhythm, melody, and harmony.  However,  their Shannon entropy reaches maximum when the loss rate exceeds 0.8. This implies that the word loss renders the ambient sound more music-like. We suspect that the deletion causes the originally continuous noise to become a combination of discrete segments, which is a feature of music. When the loss rate approaches 0.9, all non-music is transformed into music.

In Fig. 8(b), we normalized the grad CAM score into 10 equal parts, not only to calculate the proportion of 2D EEN, but also to understand the grouping structure of Fig. 7(c). We used graph theory to calculate the density, ${2L}/[N(N-1)]$ where $L,N$ denote the link and node numbers.
A link is defined if the distance is less than the  average separation between each score point and all its neighbors. We found that, although non-music  has a higher proportion than music in each scoring part, its density is lower than that of music. This suggests that non-music characteristics have a more scattered features than music.
 We also analyzed the word frequency of texts from the three periods of CPP and Japanese modern music using word cloud technology\cite{cloud}. Since CC=1 remains the most frequent word in all periods, we removed it to concentrate on the rest of words in Fig. 8(c-f). They show that words in EEN evolve throughout each period, while some are eventually terminated, just like words in natural language.

\section{Conclusion and Discussions}

 By mapping the words of  musical texts into 1 and 2 D in EEN, we discovered different regularity in composition structures of Baroque, Classical and Romantic period music. Our approach from a more scientific angle allows us to not only (1) obtain several results that are in line with current musical understandings, but also 
 (2) differentiate non-music through its higher graph density from music, and offer insights into the evolution of musical texts.
 Among the conclusion for (1), we found that (i) Baroque period is characterized by a more rigorous and ordered structure that emphasizes on repeating the same form, as exemplified by the repetition of a particular pattern in works like Fugue and Johann Pachelbel’s Canon \cite{FU,CANON},  
(ii) although Beethoven falls between the Baroque and Romantic periods, the  characteristics of his music are closer to the structure of the Baroque period, (iii) the arrangement of words for Romantic music distinctly differs from the two preceding  periods, which supports the emphasis on individualism by Romantic composers.

Preliminary results  for (2) suggest that it is promising to promote  EEN to music with non-equal temperament laws. For instance, we may use the  weights of Austronesian music as an indicator for the evolution and migration of Austronesian peoples \cite{S1,S2}, much like the role of genes \cite{Zipfs2}. Our approach has good potential to substantially advance the field of anthropology \cite{ant1}.

If 2D's EEN is the result of converting music into digitized musical text, it implies that if we can automatically generate a 2D EEN generator trained on a specific period, we could potentially use such a text-based network approach for music generation, for example, by utilizing the generated antagonism network (GAN). \cite{GAN1}. Different from directly manipulating music by audio format \cite{NM1}. In the meantime, it is recommendable to incorporate cycle-GAN \cite{ CYC_G} between 2D EEN and the Mel spectrum \cite{MEL1,MEL2} to explore how words are presented in music information.
% We set the epoch to 200, the learning rate to 0.0001, and the batch size to 64

A knowledge graph, through processes such as numerical and vector representation, i.e., Word2VEC and Harmonic mean\cite{WORD2,HARMO}, aims to deduce the semantic relationships between textual concepts and their interconnections. By leveraging entity semantics, it forms a network of knowledge. Our text network EEN, denoted as CC, can serve as a scalar for textual information. Furthermore, the text network inherently organizes contextual relationships to help us understand the semantic meaning behind music. For the increasingly active field of natural language processing in deep learning, there is a widespread lack of interpretability. Remedying this problem, our definition of words can directly correspond to the actual power spectrum.

Schopenhauer thinks that music is an embodiment of will \cite{SCP}, and how exactly emotions are expressed through music has always been a topic of debate \cite{Philosophy of Music}. By offering a basis to extract the meaning behind music in a systematic and quantitative way, we show that there may indeed be a ``language of the emotions"\cite{Philosophy of Music} that musicians have cultivated throughout history.

Financial support from the National 380 Science and Technology Council in Taiwan under Grants 381 No. 111-2112-M007-025 and No. 112-2112-M007-015 is 382 acknowledged.

{}
\end{document}